# Prospects For A New Light–Nuclei, Fission–Fusion Energy Cycle

R. V. Duncan [a*], Cuikun Lin [a], Andrew K. Gillespie [a], and John Gahl [b]

**AFFILIATIONS**

[a] *Department of Physics and Astronomy, Texas Tech University, Lubbock Texas 79409, USA*
[b] *Department of Electrical and Computer Engineering, University of Missouri, Columbia Missouri 65211, USA*
*Author to whom correspondence should be addressed: Robert.Duncan@ttu.edu

**ABSTRACT**

Future advanced nuclear rocket propulsion, and the availability of new nuclear power cycle designs, will benefit substantially from the large current investment in alternative nuclear energy that is underway today. We propose a new nuclear cycle which includes the primary fission of lithium-6, followed by secondary fusion of deuterium and tritium, and a secondary fission of lithium-7 by tritium. This cycle does not produce nuclear waste from its nuclear fuel, since all byproducts of these cascade reactions are stable, provided that the triton production during the primary reaction is fully consumed in the secondary reactions. This cycle may, however, activate surrounding technical materials from its neutron flux. This light-element nuclear fuel is readily obtained through the ongoing expansion of the lithium mining industry and electric vehicle (EV) battery recycling industries.

**I. INTRODUCTION**

As discussed by Gahl *et al.*[1] in this volume, fission fragment propulsion was proposed in 1988 by George Chapline of the Lawrence Livermore National Laboratory,[2] and further considered for use in a dusty plasma reactor design by Sheldon and Clark in 2005.[3] This fission fragment rocket design utilizes typically sub-micron sized nuclear fuel, which permits the fission fragments to escape the fuel element without depositing a substantial amount of its kinetic energy in the fuel as heat. This approach, which is only used once the rocket is well outside of the Earth's atmosphere, provides a much higher specific impulse, since it utilizes the fission fragments with a rocket exhaust velocity of about 5% of the speed of light. It also exhausts the mass of the nuclear fuel as the rocket operates. These two advantages provide much greater efficiency in nuclear propulsion over the more conventional approaches of nuclear thermal rockets and nuclear electric ion thrust rockets that are being tested for a human Mars mission today.[4]

We propose the new, light-element, fission – fusion nuclear cycle that is displayed schematically in **Figure 1**, which may be used in future nuclear rocket designs, and for the production of electrical power. In addition to the absence of fuel-related nuclear waste, this cycle is almost twice as efficient in conversion of mass into energy than conventional $^{235}$U fission reactors, where the efficiency $\varepsilon = Q / mc^2$, where $Q$ is the energy released by the reaction and $m$ is the mass of all reactants. This cycle may be operated in a sub-critical configuration and pumped with an external neutron source,[1] permitting another major advantage: in this sub-critical operational modality, the nuclear reaction cannot run-away, since when the external neutron source is off, the reaction stops. If desired, this cycle may also be used in combination with heavier element fission reactions to sustain a critical reaction.

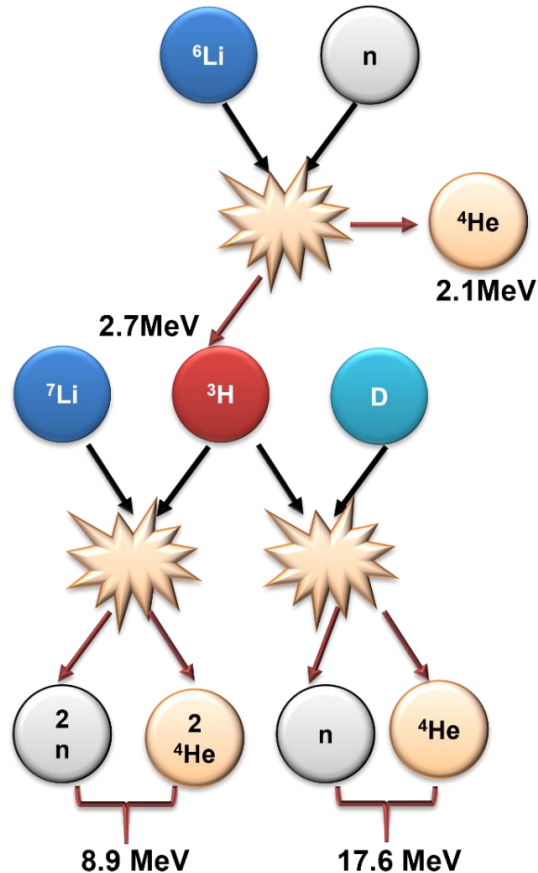

**Figure 1:** The proposed new light-element, fission – fusion hybrid fuel cycle. The triton ($^3$H) has ~70 μm range of travel in LiD, resulting in a the nearly complete consumption of the tritium that is produced in the primary [$^6$Li(n,t)$^4$He] fission cycle in the two secondary reactions, specifically the fission reaction [$^7$Li(t,2n+α)$^4$He] and the secondary fusion reaction [D(t,n)$^4$He]. The neutrons produced in the secondary reactions are energetic, especially the 14.1 MeV neutron from the D-T fusion, and they will produce collisional accelerations of the deuterium fuel and hence higher-order reactions. In this analysis, only the energetics from this primary and these two immediate secondary reactions are quantified, but the overall release of energy from this cycle will be higher.

It is desirable for the metallic deuteride particles used as fuel in this cycle to be less than a micron in diameter, so that the charged nuclear fragments from these reactions can readily escape the fuel without depositing a substantial amount of energy into the fuel particles as heat.[1–3] This permits direct conversion of the nuclear fragment energy into electricity through direct conversion, as well as the new type of nuclear fragment propulsion for rocketry while outside of the Earth's atmosphere, as discussed briefly above, and in more detail within this volume by Gahl *et al.*,[1] and by Weed, *et al.*[5]

The lithium mining and processing infrastructure necessary to support this new form of nuclear energy already exists at scale, due to the emergence of the Electric Vehicle (EV) industry. The development of energy production from this proposed cycle in association with lithium battery recycling plants, and hence in a de-centralized fashion, supports a strategy to bring the EV industry fully to scale in the future without placing unrealistic demands on our aging national electrical grid.

## II. MATERIALS AND METHODS

The primary nuclear reaction used to start this cycle described in **Figure 1** utilizes thermal neutron capture in $^6$Li, via $^6$Li(n,t)$^4$He, which has a $Q = 4.78$ MeV and a cross section of ~ 400 barns at thermal incident neutron energies, based upon NDEF calculations[6] of the cross section of this reaction as a function of energy of the incoming neutron. The resulting energetic triton from this reaction has a kinetic energy of ~ 2.7 MeV, and it may be used to induce one of two probable secondary nuclear reactions, either collisional fusion: d(t, n)$^4$He ($Q = 17.6$MeV, σ ~ 5 barns at 70 keV), or fission: 7Li(t, 2n+alpha)4He (Q = 8.8 MeV, σ = 1.2 barns at 1.8 MeV). These two secondary reactions may be made to be about equally likely, since the number density of D and Li in the LiD are the same at $5.9 \times 10^{22}$/cm$^3$, and since the peak cross-sections and reactivity of both the high-energy triton fission of $^7$Li, and the collisional fusion of the energetic triton with D, are roughly the same, yet at different energies. This number density of deuterium and lithium in LiD was calculated from the density of LiD, which is 0.883 g/cm$^3$,[7] and from its molar mass, which is 8.955 g/mole,[8] and it is typically about seven orders of magnitude larger than the typical deuteron density in a D-T fusion plasma.

The fission cross-section of $^7$Li peaks at an incident triton energy of 1.75 MeV,[6] which is about 1 MeV lower than the 2.73 MeV triton energy from the primary reaction. The peak in the D-T fusion cross-section of ~ 5 barns occurs at a much lower incident triton energy of about 70 keV.[6] But the range of the 2.7 MeV triton in lithium deuteride is only about 70 microns,[9] with many of the elastic collisions of the triton leading to energy loss along its path, until it reaches the point where the triton's energy is resonant with a peak of either the fission (with $^7$Li) or fusion (with D) cross-sections. With appropriate fuel particle size selection, all energetic tritons from the primary reactions will be involved in one of the two secondary reactions. An optimal design for this proposed new cycle will include a thin (~ 1 mm) layer of isotopically enriched $^6$Li metal closest to the incoming stream of thermal neutrons, followed by a ~ 1 cm layer of LiD particles of one micron radius. Tests will be conducted following future thermal neutron irradiations of these structures to measure the level of trapped tritium in these Li / LiD samples to infer the level of

triton production that was not involved in either of the primary or secondary reactions described above.[10]

Of these two secondary reactions, the D-T fusion reaction produces a much more energetic neutron at 14.1 MeV. This highly energetic neutron is above the (n,2n) reaction energy threshold of ~ 8 MeV in many heavy elements, such as in the stable isotopes of erbium, which may be used as a hydride to rapidly moderate these D-T fusion neutrons while also increasing the neutron number to match that resulting from the competing fission secondary reaction: $^7$Li(t, 2n + α)$^4$He. Use of erbium deuteride in this cycle advances the opportunity to sustain a fission / fusion chain reaction through neutron number amplification, which is not possible from the primary $^6$Li fission with D-T fusion alone, due to the lack of neutron number amplification. This inclusion of a heavy metal deuteride, such as $ErD_2$ or $ErD_3$, in the fuel cycle would produce some fuel-related nuclear waste. It may also be possible to sustain a fusion chain reaction through neutron collisional acceleration of the deuterons to participate in D-D fusion reactions, or possibly D-T fusion reactions if tritium were used instead of deuterium in the metal hydrides within this proposed cycle. Finally, this proposed cycle may be used in combination with conventional fission reactors to increase the reactor yield without increasing the fuel-related nuclear waste.

We note in passing that this proposed cycle in **Figure 1** produces much more energy through its primary and secondary reactions than does the p + $^{11}$B reaction that produces three alpha particles,[6,11] and which is the basis of the plasma reactors that are under development at TAE Systems.[12] The TAE approach utilizes aneutronic fusion with hydrogen-boron fuel, while this proposed LiD cycle (**Figure 1**) may be operated in a sub-critical mode using an external thermal neutron source, or it may be made to go critical in more conventional reactor designs.

## III. RESULTS AND DISCUSSIONS
### IIIa. RESULTS

To test the feasibility of this new cycle, we utilized moderated neutrons from a sealed ~2 mCi $^{252}$Cf spontaneous fission source to consecutively irradiate three columns, the first being a column of air, the second being a column of lithium metal particles, and the third being a column of lithium deuteride particles. All of the lithium used in this experiment was confirmed to contain the naturally-occurring stable isotope distribution of 7.6% $^6$Li and 92.4% $^7$Li, using a LA-ICP-MS isotopic analysis of our samples.[13] This isotopic confirmation is important, since there is already a substantial amount of $^6$Li isotopic purification going on in the current lithium supply chain today,[14] and the presence of this $^6$Li isotope is essential in order to initiate this proposed cycle from a thermal neutron source.[13] Each column was 53.3 cm long and 9.14 cm in diameter, containing a stack of three double-walled plastic containers that were partially filled. The lithium metal column contained 420 g of naturally-occurring lithium, and the LiD column contained 541 g of LiD. These masses were chosen to provide an equal 60.5 moles of each substance in each of these two columns. The third column was empty, containing only ambient air, to provide a free path for neutrons from the $^{252}$Cf source to our suite of radiation sensors at the top of the column. The sensor set at the top of each column included three neutron counters, specifically two Kromek TN15 counters and a Canberra S-NS counter, a Kromek GR1 gamma spectrometer, and a high-energy neutron spectrometer which consisted of an Elgin EJ-309 liquid scintillator and a photomultiplier tube.

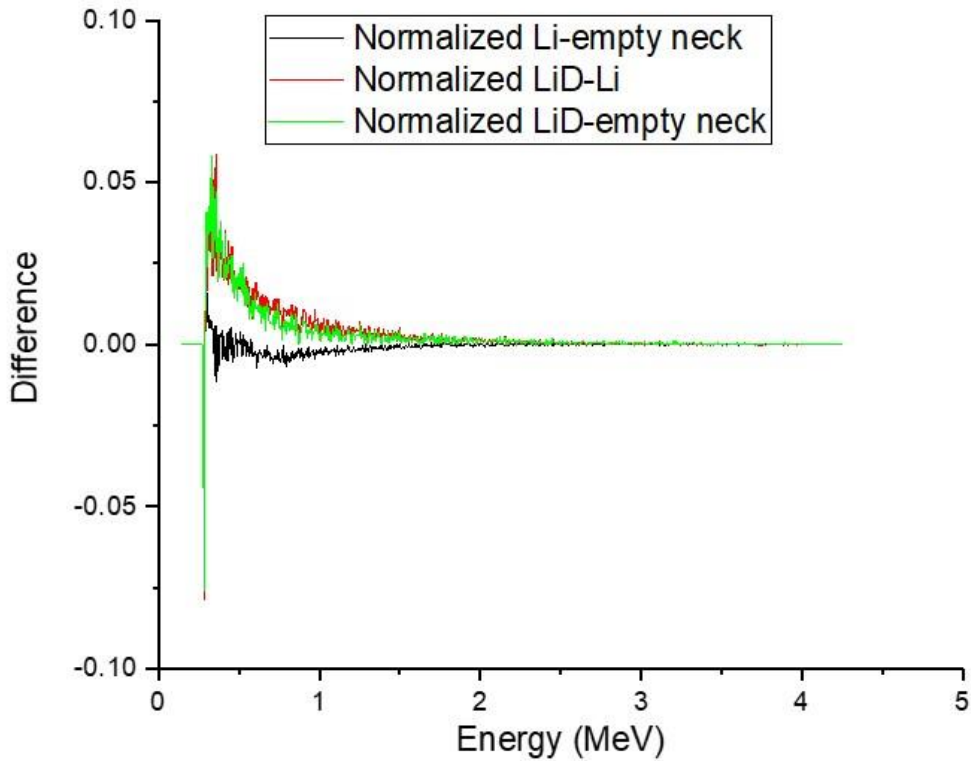

**Figure 2:** The results of the column irradiation using a 2 mCi source of $^{252}$Cf. An increased neutron energy spectrum was observed extending up to approximately 2 MeV only when the LiD column was irradiated.

The results of these measurements are displayed in **Figure 2**. The light elements moderated and attenuated the neutron flux substantially. The lithium metal column reduced the neutron count at the top column to 28% of the empty neck readings, and the LiD column reduction level was to 10%, as measured by the neutron spectrometer. Each set of measurements in **Figure 2** were normalized to their peak channel counts on the neutron scintillator spectrometer, and then differences between these normalized distributions were plotted as a function of neutron energy in **Figure 2**. Notice that the LiD column displayed a clear higher-energy tail in its neutron energy distribution than either the empty neck column or the column containing only the Li metal. This data is consistent with the secondary interaction of the energetic tritons (that resulted from $^6$Li thermal neutron capture) with deuterons in the LiD producing enough high-energy neutrons to bias the normalized neutron distribution to higher energies over the range extending up to about 2 MeV. While this initial data is indicative of triton-secondary fusion reactions, these data are preliminary, and we plan to take another series of measurements that are similar to these, but using the much higher flux (~ $10^9$ n/cm$^2$/s) thermal neutron port within the University of Missouri Research Reactor (MURR) in the near future. These future measurements will involve the irradiation of LiOD, rather than simply LiD, in order to test for collisional fusion of the energetic triton with $^{16}$O to produce $^{18}$F, which will provide a prompt 511-keV positronium annihilation gamma that subsides with the 110-minute half-life of the $^{18}$F, if the collisional fusion phenomenon is present in these samples. In these upcoming measurements, we will also experiment with the use of a thin (1-mm) layer of isotopically pure $^6$Li on the outer wall of the sample holders, in an attempt to increase the flux of energetic tritons into the LiOD sample interior.

## IIIb. COLLISION-ACCELERATED, LATTICE-CONFINED (CALC) FUSION

This proposed cycle is an example of a new class of nuclear fusion reactions that utilizes particle collisions within the solid-state lattice to induce fusion reactions within the lattice. The emergence of this new type of lattice-confined fusion reactions promises to eliminate the need for the expensive and massive equipment that is currently required to support magnetic or inertial confinement fusion.[15] The neutron collision-induced fusion also moderates the neutron energy for use in subsequent fission reactions, effectively harvesting the neutron energy that would otherwise be lost to thermal energy through more conventional moderation. This will result in a practical, solid-state approach to fusion power, which is essential to its ultimate large-scale utilization.

The density of LiD molecules ($5.9 \times 10^{22}$ atoms per cm$^3$) provides high-density confinement of the fusion (D) and fission (Li) fuels. The production of an energetic triton by $^6$Li thermal neutron capture in our proposed cycle, or alternatively by the collisional acceleration of deuteron by energetic neutrons to initiate collisional fusion reactions with stationary deuterons elsewhere in the solid-state fuel, is essential to achieve this new form of solid-state nuclear energy.

This new type of collision-induced, solid-state fusion reaction was first reported in titanium deuteride and in erbium deuteride by Steinetz, *et al.*,[16] and theoretically analyzed by Pines, *et al.*[17] Pines *et al.* analyzed the Steinetz *et al.* experiment, and predicted that the elastic neutron collisions, D(n,n)D, on average would produce 64 keV deuterons, resulting in only a 17 mb cross-section for fusion with a lab-stationary deuteron elsewhere in the sample. In contrast, the smallest cross section in our proposed LiD cycle (**Figure 1**) was 1.2 b. We suspected that this very small 17 mb cross section may be inadequate to initiate the fusion cycle that was reported in Steinetz, *et al.*, but this important suggestion of a new lattice-confined fusion mechanism deserved very close scrutiny, resulting in our experimental repetition attempts described below.

We have attempted to confirm this collision-induced, lattice-confined fusion mechanism reported within Steinetz, *et al.* using neutron irradiations of TiD$_2$ (and TiH$_2$ controls) within the MURR facilities, rather than by an accelerator-driven photo-neutron irradiations as in Steinetz, *et al*. The MURR reactor irradiated our sealed TiD$_2$ samples with $8.2 \times 10^{13}$ n/s cm$^2$ for 50 hours, and quantitative tritium measurements following extraction from the samples after irradiation were used to measure the rate of any neutron collision-induced D-D fusion in the TiD2 samples. The flux within the MURR reactor at the position of the irradiation consisted of 10% fast, and 90% thermal or epithermal neutrons. This high thermal flux of neutrons produced tritium through the well-known D(n,γ)T reaction, and any $^3$He produced by the collision-induced fusion reaction D(D,n)$^3$He would rapidly burn-up into tritium through the reaction $^3$He(n,p)T given the high thermal neutron flux, so no evidence of $^3$He production in our irradiated samples was detected using our advanced ion-cyclotron resonance mass spectrometry measurement system,[18] and none was expected. Evidence of tritium generation was measured using a 2009 Quantulus GCT 6220 scintillator. A level of $(5.4 \pm 0.4) \times 10^{12}$ tritium atoms were detected in each of the three experimental

TiD$_2$ samples. A simplified model of the samples was simulated using the Monte Carlo n-Particle nuclear transport code (MCNP6.2) for TiD$_2$ contained in a cylindrical quartz ampule. This MCNP simulation predicted that 3.9x10$^{12}$ tritium atoms would be generated in the experiment, based upon the best available characterization of the reactor's neutron spectrum and flux. We estimate that reactor variations and geometric effects could have produced up to a 25% simulation uncertainty. Any significant deviation between the experimental results and the simulated estimations might have indicated additional induced collisional fusion. However, the level of tritium production that we measured was not inconsistent with these known reactions listed above. Therefore, we are unable to confirm the existence of additional fast-neutron induced D+D collisional fusion tritium production from these data, resulting in an unsuccessful first attempt to replicate the results of Steinetz, *et al*. The uncertainty in our measurements and simulations indicate that additional replication attempts are needed before we can become definite regarding the validity of collisional-induced fusion in the Steinetz, *et al*. results.

We are attempting, once again, to replicate the Steinetz, *et al*. results, but this time we are using an accelerator-based neutron source that is similar to the original work by Steinetz, *et al*. Specifically, we are using the MURR cyclotron facility, which produces 5x10$^9$ n/s of energetic neutrons within the radiation vault while the machine operates to produce various medical radioisotopes. This method of neutron production is advantageous over the MURR reactor irradiations because it does not produce thermal or epi-thermal neutrons, as measured by Jeffries, *et al*.[19] This lack of a thermal neutron flux will greatly reduce the uncertainty of the level of collision-induced fusion that occurred in the MURR reactor irradiations. We are irradiating both TiD$_2$ experimental samples and TiH$_2$ control samples for one month of total irradiation time within this cyclotron vault. This experiment will be completed soon, providing a definitive result that either successfully or unsuccessfully replicates the Steinetz, *et al*. results.

**IIIc. MATERIALS INFRASTRUCTURE TO SUPPORT THIS PROPOSED CYCLE**

A typical electric passenger vehicle sold today contains 8 kg of lithium (about 1,150 moles of lithium)[20]. Of this, the natural abundance of $^6$Li is 7.6%, so if the lithium in a typical electric car battery is wild-type, then about 86 moles of this lithium will be the lighter, fissile isotope $^6$Li. Only 11% of the lithium in a typical passenger car battery (to include all of the $^6$Li within this battery), supplying our proposed new nuclear fission / fusion cycle described herein, will produce over 4.2x10$^7$ kWh of energy, which is equivalent to the energy required to fully charge 184,000 such car batteries to their full 75 kWh charge, resulting in over 45+ million passenger vehicle driving miles, assuming a 35% energy conversion efficiency from heat to electricity. This is equivalent to the typical lifetime driving distance of 450+ such vehicles. So, the lithium-6, and about half this amount of lithium-7, or a total of 86 + 43 = 129 moles of lithium = 0.817 kg, is all that is required from the single recycling of one single car battery to supply all of this energy. This implies that the entire U.S.A. automotive industry may re-tooled to 'all electric', and may be fully energy-supported, by just 0.25% of the spent batteries being recycled in this manner, using this new nuclear cycle. We note that this approach is quite feasible, since the mining, recycling, and fabricating of lithium already exists at scale. This energy production may be done in close proximity to existing lithium reprocessing capabilities.

The world's known lithium reserves are currently estimated at $2.3 \times 10^{10}$ kg,[21] and much more of these reserves may be mined in the future. If all of the currently known world reserves of lithium were consumed using this proposed cycle, most significantly if all of the world's known $^6$Li were consumed, then almost a millennium of the world's current energy needs (now $5.8 \times 10^{20}$ J per year) could be met.[22] Once again, the primary advantage of this light-element, fission – fusion cycle is that no fuel-related nuclear waste would be produced.

## IV. ACKNOWLEDGEMENTS

We are grateful to Ian Jones, who measured the level of tritium in our samples, and to Matthew Looney (Texas Tech) for their suggestions and their efforts as the Radiation Safety Officers involved in this effort. We thank Prof. John Brockman from MURR for his advice and recommendations associated with the radiations reported herein.

## V. DATA AVAILABILITY

The data and methods recorded and utilized in this study can be found at https://www.depts.ttu.edu/phas/cees/, and through the Information Technology Division of Texas Tech University. Other data that support the findings of this study are available from the corresponding author upon reasonable request.

## VI. REFERENCES

1. J. Gahl, , A. Gillespie, R.V. Duncan, C. Lin. "The Fission Fragment Rocket Engine for Mars Fast Transit," Frontiers in Space Technologies. Volume 4 (2023). doi.org/10.3389/frspt.2023.1191300  ISSN=2673-5075

2. G. Chapline, "Fission Fragment Rocket Concept," Nuclear Instruments and Methods in Physics Research Section A 271, 1 (1988).

3. R.L. Clark and R. B. Sheldon, "Dusty Plasma Based Fission Fragment Nuclear Reactor," in American Institute of Aeronautics and Astronautics, Inc., Tucson, AZ (2005).

4. The Guardian (2023), NASA to test nuclear rockets that could fly astronauts to Mars in record time. https://www.theguardian.com/science/2023/jan/24/nasa-mars-trip-nuclear-rocket [Accessed March 15, 2023].

5. R. Weed, R.V. Duncan, M. Horsley, and G. Chapline. "Radiation Characteristics of an Aerogel-Supported Fission Fragment Rocket Engine for Crewed Interplanetary Missions," Frontiers in Space Technologies. Volume 4 (2023). doi.org/10.3389/frspt.2023.1197347  ISSN=2673-5075

6. G.M. Hale, ENDF/B-VIII.0: LI-6(n,T)HE-4, MT105 QM=4.78365e+6, Nucl. Data Sheets 107(2006)2931.

7. E. Staritzky and D.I. Walker, Anal. Chem. 1956, 28, 6, 1055. https://doi.org/10.1021/ac60114a043.

8. E.M. Baum et al., Nuclides and Isotopes, Chart of the Nuclides, Seventeenth Edition (Bechtel Marine Propulsion Corporation, ISBN 978-0-98443653-0-2, 2010)

9. H.V. Ruiz, International Journal of Radiation Applications and Instrumentation, Part A: Applied Radiation and Isotopes 1988, 39 (1), 31-39. DOI: https://doi.org/10.1016/0883-2889(88)90089-5.

10. Texas Tech University Center for Emerging Energy Sciences (2022), Measurement of tritium from metallic hydride samples. https://www.depts.ttu.edu/phas/cees/Ref/Q_Tritium.pdf  [Accessed March 15, 2023].

11. W.M. Nevins and R. Swain, Nuclear Fusion 40, 865 (2000).

12. TAE (2023), First measurements of hydrogen-boron fusion in a magnetically confined fusion plasma.   https://tae.com/first-measurements-of-hydrogen-boron-fusion-in-a-magnetically-confined-fusion-plasma/ [Accessed March 15, 2023].

13. R.I. Gabitov et al., (2011), In situ δ7Li, Li/Ca, and Mg/Ca analyses of synthetic aragonites, Geochem. Geophys. Geosyst., 12, Q03001, doi:10.1029/2010GC003322

14. E.A. Symons, "Lithium Isotope Separation A Review of Possible Techniques," Report No. CFFTP-G-85036, Cross Ref. Report No. AECL-8708

15. B. Baramsa, et al., "NASA's New Shortcut to Fusion Power", IEEE Spectrum, February 27, 2022.

16. B. Steinetz, et al., NASA TP-20205001616, and Phys. Rev. C 101, 044610 (2020).

17. V. Pines, et al., NASA TP-20205001617, and Phys. Rev. C 101, 044609 (2020)

18. R.P. Thorn et al., "A quantitative light-isotope measurement system for climate and energy applications," International Journal of Mass Spectrometry, Volume 464, 2021, 116574, ISSN 1387-3806, doi:10.1016/j.ijms.2021.116574.

19. B.D. Jeffries, et al., Appl. Rad. Isot. 154 108892 (2019)

20. World Economic Forum (2022), Lithium batteries. https://www.weforum.org/agenda/2022/07/electric-vehicles-world-enough-lithium-resources/ [Accessed March 15, 2023].

21. Statista (2023). Reserves of lithium worldwide as of 2022, by country. https://www.statista.com/statistics/268790/countries-with-the-largest-lithium-reserves-worldwide/ [Accessed March 15, 2023].

22. The World Counts (2023). Global Energy Consumption. https://www.theworldcounts.com/challenges/energy/global-energy-consumption [Accessed March 15, 2023].